\documentstyle[twoside,ncs,psfig,fleqn,espcrc2]{article}
\def\bnue{\hbox{$\bar\nu_e$ }}  
\def\bnum{\hbox{$\bar\nu_\mu$ }}  
  
\newcommand {\ignore}[1]{}

\newcommand{\bc}{\begin{center}}
\newcommand{\ec}{\end{center}}

\def\ifmath#1{\relax\ifmmode #1\else $#1$\fi}
\def\half{\ifmath{{\textstyle{1 \over 2}}}}

\def\3quarter{{\textstyle{3 \over 4}}}

\def\ra{\rightarrow}

\def\lf{\leaders\hbox to 1em{\hss.\hss}\hfill}
\def\e6{$E(6)$}
\def\10{$SO(10)$}
\def\21{$SU(2) \otimes U(1) $}
\def\lr{$SU(2)_L \otimes SU(2)_R \otimes U(1)$}
\def\422{$SU(4) \otimes SU(2) \otimes SU(2)$}
\def\321{$SU(3) \otimes SU(2) \otimes U(1)$}
\def\ne{\hbox{$\nu_e$ }}
\def\nm{\hbox{$\nu_\mu$ }}
\def\nt{\hbox{$\nu_\tau$ }}
\def\ns{\hbox{$\nu_{s}$ }}

\def\mnt{\hbox{$m_{\nu_\tau}$ }}

\def\neus{\hbox{neutrinos }}

\def\neu{\hbox{neutrino }}

\def\eq#1{{eq. (\ref{#1})}}

\def\fig#1{{Fig. (\ref{#1})}}

\def\VEV#1{\left\langle #1\right\rangle}
\let\vev\VEV

\def\lsim{\raise0.3ex\hbox{$\;<$\kern-0.75em\raise-1.1ex\hbox{$\sim\;$}}}
\def\gsim{\raise0.3ex\hbox{$\;>$\kern-0.75em\raise-1.1ex\hbox{$\sim\;$}}}
\def\half{{1\over 2}}
\def\beq{\begin{equation}}
\def\eeq{\end{equation}}
\def\bef{\begin{figure}}
\def\eef{\end{figure}}
\def\bet{\begin{table}}
\def\eet{\end{table}}
\def\bea{\begin{eqnarray}}
\def\ba{\begin{array}}
\def\ea{\end{array}}
\def\bi{\begin{itemize}}
\def\ei{\end{itemize}}
\def\ben{\begin{enumerate}}
\def\een{\end{enumerate}}
\def\ra{\rightarrow}

\def\eea{\end{eqnarray}}
\def\apj#1#2#3{          {\it Astrophys. J. }{\bf #1} (19#2) #3}

\def\ib#1#2#3{           {\it ibid. }{\bf #1} (19#2) #3}
\def\nat#1#2#3{          {\it Nature }{\bf #1} (19#2) #3}
\def\nps#1#2#3{        {\it Nucl. Phys. B (Proc. Suppl.) }{\bf #1} (19#2) #3} 
\def\np#1#2#3{           {\it Nucl. Phys. }{\bf #1} (19#2) #3}
\def\pl#1#2#3{           {\it Phys. Lett. }{\bf #1} (19#2) #3}
\def\pr#1#2#3{           {\it Phys. Rev. }{\bf #1} (19#2) #3}
\def\prep#1#2#3{         {\it Phys. Rep. }{\bf #1} (19#2) #3}
\def\prl#1#2#3{          {\it Phys. Rev. Lett. }{\bf #1} (19#2) #3}

\def\rmp#1#2#3{          {\it Rev. Mod. Phys. }{\bf #1} (19#2) #3}
\def\zp#1#2#3{           {\it Zeit. fur Physik }{\bf #1} (19#2) #3}

\def\n.c.#1#2#3{         {\it Nuovo Cim. }{\bf #1} (19#2) #3}
\def\r.n.c.#1#2#3{       {\it Riv. del Nuovo Cim. }{\bf #1} (19#2) #3}
\def\sjnp#1#2#3{         {\it Sov. J. Nucl. Phys. }{\bf #1} (19#2) #3}

\def\jetp#1#2#3{         {\it JETP }{\bf #1} (19#2) #3}
\def\mpl#1#2#3{          {\it Mod. Phys. Lett. }{\bf #1} (19#2) #3}

\def\ppnp#1#2#3{           {\it Prog. Part. Nucl. Phys. }{\bf #1} (19#2) #3}

\def\pc{private communication}

\def\ip{in preparation}

\newcommand{\AmS}{{\protect\the\textfont2
  A\kern-.1667em\lower.5ex\hbox{M}\kern-.125emS}}
\title{Neutrinos Properties Beyond the Standard Model}
\author{Jos\'e W. F. Valle\address{Instituto de F\'{\i}sica Corpuscular 
- C.S.I.C.\\Departament de F\'{\i}sica Te\`orica, Universitat de 
Val\`encia\\46100 Burjassot, Val\`encia, Spain}
\thanks{ Supported by DGICYT grant PB95-1077 and in part by EEC under
the TMR contract ERBFMRX-CT96-0090. I thank the organizers for their
hospitality, as well as Plamen Krastev and Eligio Lisi for providing 
me the files of figures 9 and 10.  }}
\begin{document}
\begin{abstract}
The present observational status of neutrino physics is sketched, with
emphasis on the hints that follow from solar and atmospheric neutrino
observations, as well as dark matter. I also briefly review the ways
to account for the observed anomalies and some of their implications.
\end{abstract}
\maketitle
\section{INTRODUCTION}
\vskip .2cm

One of the biggest drawbacks of the Standard Model (SM) is that the
masslessness of neutrinos is not dictated by an underlying {\sl grand
principle}, such as that of gauge invariance in the case of the
photon: the SM simply postulates that neutrinos are massless and, as a
result, all their properties are trivial, e.g. magnetic and transition
moments are zero, etc.  Massless neutrinos would be exceptional
particles, since no other such fermions exist. If massive, neutrinos
would present another puzzle, of why are their masses so much smaller
than those of the charged fermions. The fact that neutrinos are the
only electrically neutral elementary fermions may hold the key to the
answer, namely neutrinos could be Majorana fermions, the most
fundamental ones. In this case the suppression of their mass could be
associated to lepton number conservation, as actually happens in many
extensions of the SM.

From the observational point of view non-zero neutrino masses now seem
required in order to account for the data on solar and atmospheric
neutrinos, as well as the (hot) dark matter in the universe.
Detecting neutrino masses is one of the most outstanding challenges in
particle physics, with far-reaching implications also for the
understanding of fundamental issues in astrophysics and cosmology.
Though very difficult, future experiments could shed light on the
issue of neutrino masses and the conservation of lepton number.  One
interesting aspect of many models where neutrinos have non-vanishing
masses is that they lead to effects that could be experimentally
tested. Before over-viewing the present observational limits and hints
in favour of massive neutrinos, let us make a few general remarks
about the theoretical models.

\section{THEORETICAL MODELS}
\vskip .2cm

One of the most attractive approaches to generate neutrino masses is
from unification. Indeed, in trying to understand the origin of parity
violation in the weak interaction by ascribing it to a spontaneous
breaking phenomenon, in the same way as the W and Z acquire their
masses in the SM, one arrives at the so-called left-right symmetric
extensions such as \lr \cite{LR}, \422 \cite{PS} or \10 \cite{GRS}, in
some of which the masses of the light neutrinos are obtained by
diagonalizing the following mass matrix in the basis $\nu,\nu^c$
\begin{equation}
\left[\matrix{
 0 & D \cr
 D^T & M_R }\right] 
\label{SS} 
\end{equation} 
where $D = h_D \vev{H} /\sqrt2$ is the standard \21 breaking Dirac
mass term and $M_R = M_R^T$ is the isosinglet Majorana mass. In the
seesaw approximation, one finds
\beq M_{eff} = - D M_R^{-1} D^T \:.  
\label{SEESAW} 
\eeq 
In general, however, this matrix also contains a $\nu\nu$ term
\cite{2227} whose size is expected to be also suppressed by the
left-right breaking scale. As a result one is able to explain
naturally the relative smallness of \neu masses.  Even though it is
natural to expect $M_R$ to be large, its magnitude heavily depends on
the model. As a result one can not make any real prediction for the
corresponding light neutrino masses that are generated through the
seesaw mechanism. In fact this freedom has been exploited in model
building in order to account for an almost degenerate \neu mass
spectrum \cite{DEG}.

Although very attractive, unification is by no means the only way to
generate neutrino masses. There is a large diversity of other possible
schemes which do not require any new large mass scale. For example, it
is possible to start from an extension of the lepton sector of the \21
theory by adding a set of $two$ 2-component isosinglet neutral
fermions, denoted ${\nu^c}_i$ and $S_i$, to each generation. In this
case there is an exact L symmetry that keeps neutrinos strictly
massless, as in the SM. The conservation of total lepton number leads
to the following form for the neutral mass matrix (in the basis $\nu,
\nu^c, S$)
\begin{equation}
\left[\matrix{
  0 & D & 0 \cr
  D^T & 0 & M \cr
  0 & M^T & 0 }\right] \label{MAT} \end{equation} This form has also
been suggested in various theoretical models \cite{WYLER}, including
many of the superstring inspired models.  In the latter case the zeros
of \eq{MAT} naturally arise due to the absence of Higgs fields to
provide the usual Majorana mass terms, needed in the seesaw model
\cite{SST}. Clearly, one can easily introduce non-zero masses in this
model through a $\mu S S$ term that could be proportional to the VEV
of a singlet field $\sigma$ \cite{CON}. In contrast to the seesaw
scheme, the \neu masses are directly proportional to
$\VEV{\sigma}$. This model provides a conceptually simple and
phenomenologically rich extension of the Standard Model, which brings
in the possibility that a wide range of new phenomena be
sizeable. These have to do with neutrino mixing, universality, flavour
and CP violation in the lepton sector \cite{BER,CP}, as well as direct
effects associated with Neutral Heavy Lepton (NHL) production at high
energy colliders \cite{CERN}.  A remarkable feature of this model is
the possibility of non-trivial neutrino mixing despite the fact that
neutrinos are strictly massless. This tree-level effect leads to a new
type of resonant \neu conversion mechanism that could play an
important role in supernovae \cite{massless0,massless}.  Moreover,
there are loop-induced lepton flavour and CP non-conservation effects
whose rates are precisely calculable \cite{BER,CP,3E}. I repeat that
this is remarkable due to the fact that physical light neutrinos are
massless, as in the standard model. This feature is the same as what
happens in the supersymmetric mechanism of flavour violation
\cite{Hall}. Indeed, in the simplest case of SU(5) supergravity
unification, there are flavour violating processes, like $\mu \ra e
\gamma$, despite the fact that in SU(5) neutrinos are protected by B-L
and remain massless. The supersymmetric mechanism and that of \eq{MAT}
differ in that the lepton flavour violating (LFV) processes are
induced in one case by NHL loops, while in supersymmetry they are
induced by scalar boson loops.  In both cases the particles in the
loops have masses at the weak scale, leading to branching ratios
\cite{BER,CP,3E} \cite{SUSYLFV,SUSYLFV2} that are sizeable enough to
be of experimental interest \cite{ETAU,TTTAU,opallfv}.

\begin{figure}[t]
\centerline{\protect\hbox{\psfig{file=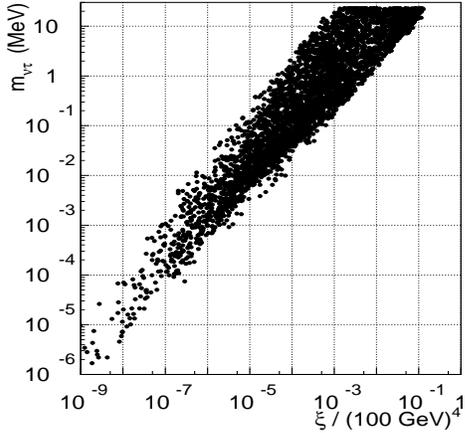,height=6cm,width=7cm}}}
\vglue -0.5cm
\caption{Tau neutrino mass versus $\epsilon_3$ }
\vglue -0.5cm
\label{mnt_xi_ev}
\end{figure}

Supersymmetry with broken R-parity also provides a nice mechanism for
the origin of neutrino mass \cite{epsrad,RPothers}. For example, in a
model where R-parity is broken by a bilinear term in the
superpotential \cite{epsrad} the tau neutrino $\nu_{\tau}$ acquires a
mass, due to the mixing between \neus and neutralinos given in the
matrix
\begin{equation}
\left[\matrix{
M_1 & 0  & -\half g'v_1 & \half g'v_2 & -\half g'v_3 \cr
0   & M_2 & \half g v_1 & -\half g v_2 & \half g v_3 \cr
-\half g'v_1 & \half g v_1 & 0 & -\mu & 0 \cr
\half g'v_2 & -\half g v_2 & -\mu & 0 & \epsilon_3 \cr
-\half g'v_3 & \half g v_3 & 0 & \epsilon_3 & 0 
}\right]
\label{eq:NeutMassMat}
\end{equation}
This mixing is proportional to the R-parity and lepton-number
violating parameters $\epsilon_3$ and $v_3$. In the simplest unified
supergravity model the $\epsilon_3$ and the $v_3$ are related
\cite{epsrad}. They lead to a non-zero Majorana $\nu_{\tau}$ mass,
which depends quadratically on an effective parameter $\xi$ defined as
$\xi \equiv (\epsilon_3 v_1 + \mu v_3)^2$. It is important to notice
that the neutrino mass generated through R-parity violation in this
model is not necessarily large, even though its implications can be
observable.  In \fig{mnt_xi_ev} we display the allowed values of
$m_{\nu_{\tau}}$ (in the tree level approximation).
As can be seen from the figure the $m_{\nu_{\tau}}$ values can cover a
very wide range, up to values in the MeV range, comparable to the
present LEP limit \cite{eps95}.  The latter places a limit on the
value of $\xi$.  Notice that \ne and \nm remain massless in this
approximation. They get masses either from radiative corrections
\cite{RPnuloops} or by mixing with singlets in models with spontaneous
breaking of R-parity \cite{Romao92}.

\begin{figure}[t]
\psfig{file=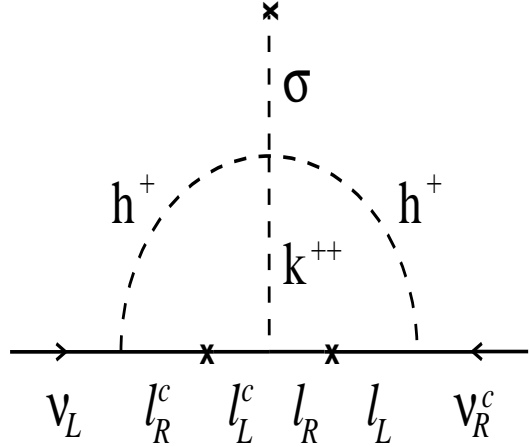,height=6cm,width=7cm}
\vglue -0.5cm
\caption{Two-loop-induced Neutrino Mass. }
\vglue -.5cm
\label{2loop}
\end{figure}

There is also a large variety of {\sl radiative} schemes to generate
\neu masses. The prototype models of this type are the Zee model and
the model suggested by Babu \cite{zee.Babu88}. In these models lepton
number is explicitly broken, but it is easy to realize them with
spontaneous breaking of lepton number. For example in the version
suggested in ref. \cite{ewbaryo} the neutrino mass arises from the
diagram shown in \fig{2loop}.

Other than the seesaw scheme, none of the above models requires a
large mass scale. In all of them one can implement the spontaneous
violation of the global lepton number symmetry leading to \neu masses
that scale {\sl directly} proportional to the lepton-number scale or
some positive power of it, in contrast to the original Majoron model
\cite{CMP}. Such low-scale models are very attractive and lead to a
richer phenomenology, as the extra particles required have masses at
scales that could be accessible to present experiments. One remarkable
example is the possibility invisibly decaying Higgs bosons
\cite{JoshipuraValle92}.

The above discussion should suffice to illustrate the enormous freedom
and wealth of phenomenological possibilities in the neutrino
sector. These reach well beyond the realm of conventional neutrino
experiments, including also signatures that can be probed, though
indirectly, at high energy accelerators.  An optimist would regard as
very exciting the fact that the neutrino sector may hold so many
experimental possibilities, while a pessimist would be discouraged by
the fact that one does not know the relevant scale responsible for
neutrino mass, nor the underlying mechanism. Last but not least, one
lacks a theory for the Yukawa couplings. As a consequence \neu masses
are not predicted and it is up to observation to search for any
possible clue.  Given the theoretical uncertainties in predicting \neu
masses from first principles, one must turn to observation. Here the
information comes from laboratory, astrophysics and cosmology.

\section{OBSERVATIONAL LIMITS ON NEUTRINO MASSES AND MIXINGS}
\vskip .2cm

\subsection{Laboratory Limits }
\vskip .2cm

The best limits on the neutrino masses can be summarized as
\cite{PDG96}:
\beq
\label{1}
m_{\nu_e} 	\lsim 5 \: \mbox{eV}, \:\:\:\:\:
m_{\nu_\mu}	\lsim 170 \: \mbox{keV}, \:\:\:\:\:
m_{\nu_\tau}	\lsim 18 \: \mbox{MeV}
\eeq
These are the most model-independent of the laboratory limits on \neu
mass, as they follow purely from kinematics.  The limit on the \ne
mass comes from beta decay, that on the \nm mass comes from PSI (90 \%
C.L.) \cite{psi}, with further improvement limited by the uncertainty
in the $\pi^-$ mass.  On the other hand, the best \nt mass limit now
comes from high energy LEP experiments \cite{eps95} and may be
substantially improved at a future tau-charm factory \cite{jj}. In
connection with tritium beta decay limit \cite{Lobashev} even though
the negative $m^2$ value has now been clarified, there are still
un-understood features in the spectrum, probably of instrumental
origin. Further results from the Mainz experiment are awaited.

Additional limits on neutrino masses follow from the non-observation
of neutrino oscillations.  The most sensitive searches have been
performed at reactors \cite{reactor} (\bnue - $\nu_x$ oscillations);
at meson factories (KARMEN \cite{karmen}, LSND \cite{lsnd}) and at
high-energy accelerators (experiments E531 and E776 \cite{E531.E776}).
A search for \nm to \ne oscillations has now been reported by the LSND
collaboration using \nm from $\pi^+$ decay in flight
\cite{lsndflight}. An excess in the number of beam-related events from
the $C(\nu_e,e^-)X$ inclusive reaction is observed. The excess cannot
be explained by normal \ne contamination in the beam at a confidence
level greater than 99\%.  If interpreted as an oscillation signal, the
observed oscillation probability of $(2.6 \pm 1.0 \pm 0.5) \times
10^{-3}$ is consistent with the previously reported \bnum to \bnue
oscillation evidence from LSND. Another recent result comes from NOMAD
and rules out part of the LSND region. The future lies in searches for
oscillations using accelerator beams directed to far-out underground
detectors, with very good prospects for the long-baseline experiments
proposed at KEK, CERN and Fermilab.

If neutrinos are of Majorana type a new form of nuclear double beta
decay would take place in which no neutrinos are emitted in the final
state, i.e. the process by which an $(A,Z-2)$ nucleus decays to $(A,Z)
+ 2 \ e^-$. In such process one would have a virtual exchange of
Majorana neutrinos. Unlike ordinary double beta decay, the
neutrino-less process violates lepton number and its existence would
indicate the Majorana nature of neutrinos.  Because of the phase space
advantage, this process is a very sensitive tool to probe into the
nature of neutrinos.

\begin{figure}[t]
\psfig{file=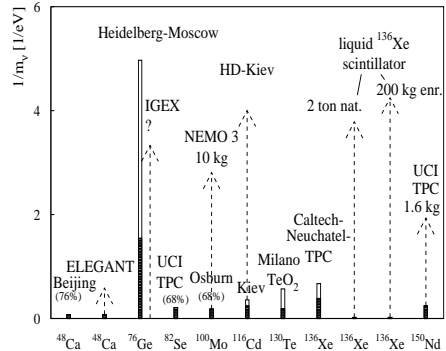,height=6cm,width=7cm,angle=-90}
\vglue -1cm
\caption{Sensitivity of ${\beta \beta}_{0\nu}$ experiments. }
\label{betabetafut}
\vglue -.5cm
\end{figure}

Present data place an important limit on a weighted average \neu mass
parameter $\VEV{m} \lsim 1 - 2$ eV. The present experimental situation
as well as future prospects is illustrated in \fig{betabetafut}, taken from
ref. \cite{Klapdor}.  
Note that this bound depends to some extent on the relevant nuclear
matrix elements characterising this process \cite{haxtongranada}.  The
parameter $\VEV{m}$ involves both neutrino masses and mixings. Thus,
although rather stringent, this limit may allow very large \neu
masses, as there may be strong cancellations between different
neutrino types. This may happen automatically in the presence of
suitable symmetries. For example, the decay vanishes if the
intermediate neutrinos are Dirac-type, as a result of the
corresponding lepton number symmetry \cite{QDN}.

Neutrino-less double beta decay has a great conceptual importance. It
has been shown \cite{BOX} that in a gauge theory of the weak
interactions a non-vanishing ${\beta \beta}_{0\nu}$ decay rate
requires \neus to be Majorana particles, {\sl irrespective of which
mechanism} induces it. This is important since in a gauge theory
neutrino-less double beta decay may be induced in other ways,
e.g. via scalar boson exchange.

\subsection{Limits from Cosmology }
\vskip .2cm

There are a variety of cosmological arguments that give 
information on neutrino parameters. In what follows I briefly 
consider the critical density and the primordial Nucleosynthesis 
arguments.

\subsubsection{The Cosmological Density Limit }
\vskip .2cm

The oldest cosmological bound on neutrino masses follows from avoiding
the overabundance of relic neutrinos \cite{KT} 
\beq 
\label{RHO1}
\sum m_{\nu_i} \lsim 92 \: \Omega_{\nu} h^2 \: eV\:, 
\eeq 
where $\Omega_{\nu} h^2 \leq 1$ and the sum runs over all species of
isodoublet neutrinos with mass less than $O(1 \: MeV)$. Here
$\Omega_{\nu}=\rho_{\nu}/\rho_c$, where $\rho_{\nu}$ is the neutrino
contribution to the total density and $\rho_c$ is the critical
density.  The factor $h^2$ measures the uncertainty in the present
value of the Hubble parameter, $0.4 \leq h \leq 1$, and $\Omega_{\nu}
h^2$ is smaller than 1.  For the $\nu_{\mu}$ and $\nu_{\tau}$ this
bound is much more stringent than the laboratory limits \eq{1}.

Apart from the experimental interest \cite{jj}, an MeV tau neutrino
also seems interesting from the point of view of structure formation
\cite{ma1}. Moreover, it is theoretically viable as the constraint in
\eq{RHO1} holds only if \neus are stable on the relevant cosmological
time scale. In models with spontaneous violation of total lepton
number \cite{CMP} there are new interactions of neutrinos with the
majorons which may cause neutrinos to decay into a lighter \neu plus a
majoron, for example \cite{fae},
\beq
\label{NUJ}
\nu_\tau \ra \nu_\mu + J \:\: .
\eeq
or have sizeable annihilations to these majorons,
\beq
\label{JJ}
\nu_\tau + \nu_\tau \ra J + J \:\: .
\eeq

The possible existence of fast decay and/or annihilation channels
could eliminate relic neutrinos and therefore allow them to have
higher masses, as long as the lifetime is short enough to allow for an
adequate red-shift of the heavy neutrino decay products. These 2-body
decays can be much faster than the visible modes, such as radiative
decays of the type $\nu' \ra \nu + \gamma$. Moreover, the Majoron
decays are almost unconstrained by astrophysics and cosmology (for 
a detailed discussion see ref. \cite{KT}).

A general method to determine the Majoron emission decay rates of
neutrinos was first given in ref. \cite{774}. The resulting decay
rates are rather model-dependent and will not be discussed here.
Explicit neutrino decay lifetime estimates are given in
ref. \cite{Romao92,fae,V}.  The conclusion is that there are many ways
to make neutrinos sufficiently short-lived and that all mass values
consistent with laboratory experiments are cosmologically acceptable.

\subsubsection{The Nucleosynthesis Limit}
\vskip .2cm

There are stronger limits on neutrino lifetimes or annihilation cross
sections arising from cosmological nucleosynthesis. Recent data on the
primordial deuterium abundance \cite{dhigh,dlow} have stimulated a lot
of work on the subject \cite{cris,ncris,sarkar}.  If a massive \nt is
stable on the nucleosynthesis time scale, ($\nu_\tau$ lifetime longer
than $\sim 100$ sec), it can lead to an excessive amount of primordial
helium due to their large contribution to the total energy
density. This bound can be expressed through an effective number of
massless neutrino species ($N_\nu$). If $N_\nu < 3.4-3.6$, one can
rule out $\nu_\tau$ masses above 0.5 MeV \cite{KTCS91,DI93}.  If we
take $N_\nu < 4$ the \mnt limit loosens accordingly. However it has
recently been argued that non-equilibrium effects from the light
neutrinos arising from the annihilations of the heavy \nt's make the
constraint a bit stronger in the large \mnt region \cite{noneq}. In
practice, all $\nu_\tau$ masses on the few MeV range are ruled out.
One can show, however that in the presence of new \nt annihilation
channels the nucleosynthesis \mnt bound is substantially weakened or
eliminated \cite{DPRV}.  Fig. 4 gives the effective number of massless
neutrinos equivalent to the contribution of a massive \nt Majoron
model with different values of the coupling $g$ between $\nu_\tau$'s
and $J$'s, expressed in units of $10^{-5}$. For comparison, the dashed
line corresponds to the SM $g=0$ case. One sees that for a fixed
$N_\nu^{max}$, a wide range of tau neutrino masses is allowed for
large enough values of $g$. No \nt masses below the LEP limit can be
ruled out, as long as $g$ exceeds a few times $10^{-4}$.

\begin{figure}[t]
\psfig{file=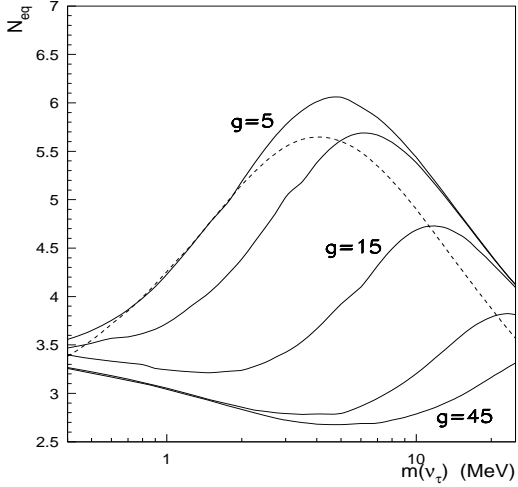,height=6.5cm,width=7cm}
\vglue -.5cm 
\caption{A heavy \nt annihilating to majorons can lower
the equivalent massless-neutrino number in nucleosynthesis.}  
\vglue -.5cm 
\label{neq} 
\end{figure}
\begin{figure}[t]
\centerline{
\psfig{file=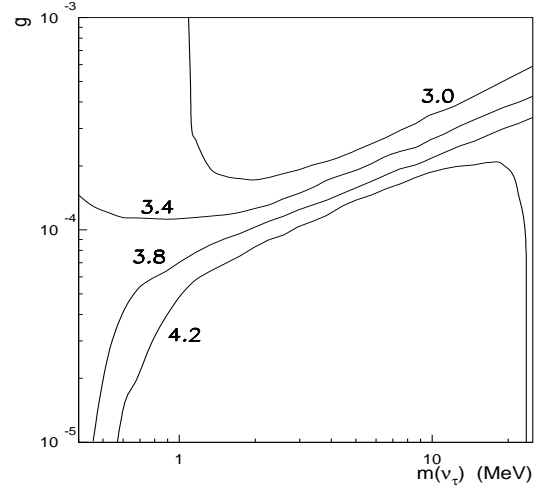,height=6.5cm,width=7cm}}
\vglue -1cm 
\caption{The region above each curve is allowed for the
corresponding $N_{eq}^{max}$.}  
\vglue -.5cm 
\label{neffmg}
\end{figure}

 
One can express the above results in the $m_{\nu_\tau}-g$ plane, as
shown in figure \ref{neffmg}.  
One sees that the constraints on the mass of a Majorana $\nu_\tau$
from primordial nucleosynthesis can be substantially relaxed if
annihilations $\nu_\tau \bar{\nu}_\tau \leftrightarrow JJ$ are
present. Moreover the required values of $g(m_{\nu_\tau})$ are
reasonable in many majoron models \cite{fae,DPRV,MASIpot3}.  Similar
depletion in massive \nt relic abundance also happens if the \nt is
unstable on the nucleosynthesis time scale \cite{unstable} as will
happen in many Majoron models.

\subsection{Limits from Astrophysics  }
\vskip .2cm

There are a variety of limits on neutrino parameters that follow from
astrophysics, e.g. from the SN1987A observations, as well as
from supernova theory, including supernova dynamics \cite{Raffelt} and
from nucleosynthesis in supernovae \cite{qian}. Here I briefly discuss
three recent examples of how supernova physics constrains neutrino
parameters.

\begin{figure}[t]
\centerline{\protect\hbox{
\psfig{file=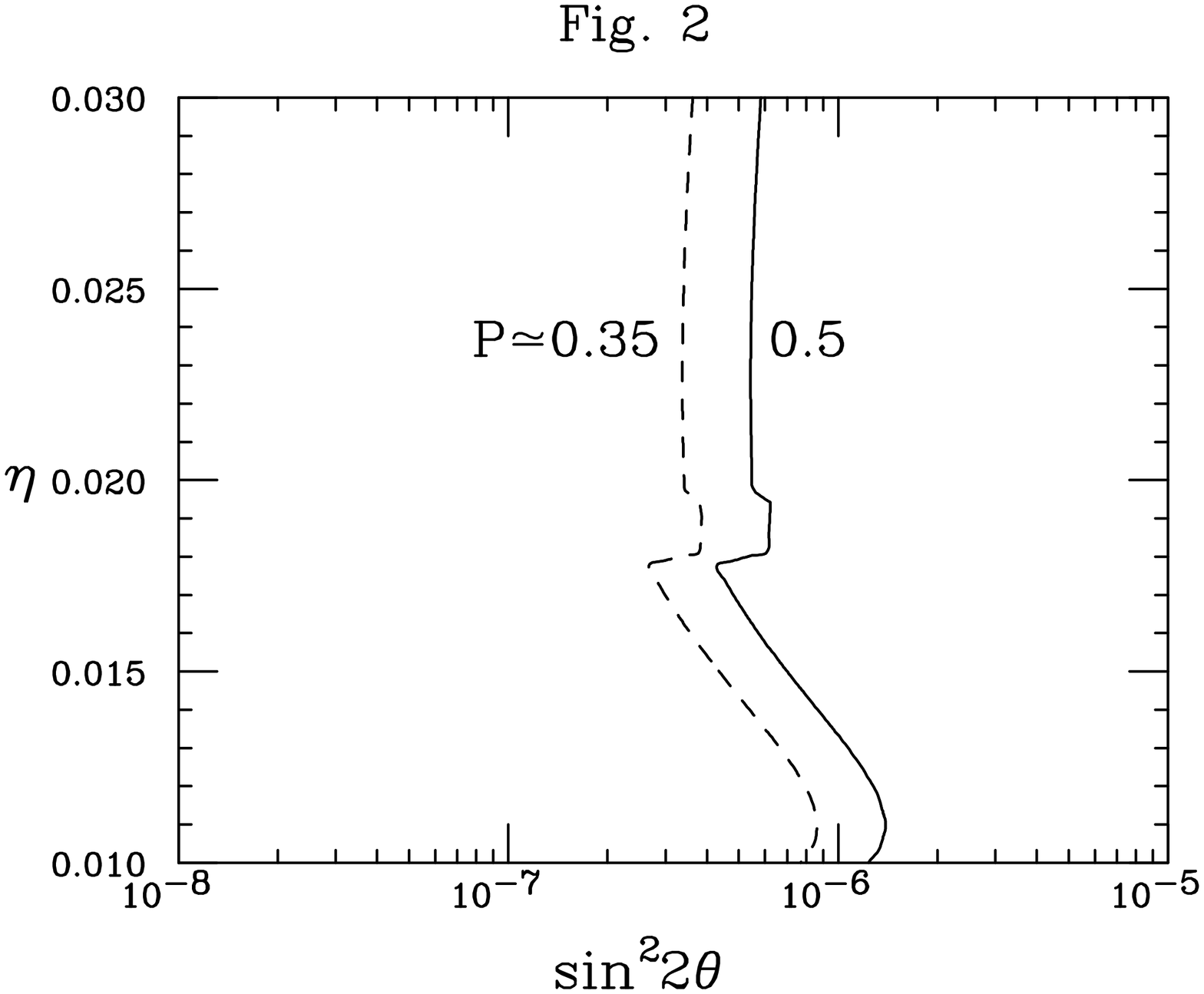,height=7cm,width=7cm,angle=90}
}}
\vglue -7.3cm
\hglue 2.7cm
\psfig{file=cover2.ps,height=0.7cm,width=3cm,angle=90}
\vglue 5.3cm
\caption{SN1987A bounds on massless neutrino mixing. }
\label{SN87}
\vglue -1cm
\end{figure}

It has been noted a long time ago that, in some circumstances, {\sl
massless} neutrinos may be {\sl mixed} in the leptonic charged current
\cite{massless0}. Conventional neutrino oscillation searches in vacuo
are insensitive to this mixing. However,  such neutrinos may
resonantly convert in the dense medium of a supernova
\cite{massless0,massless}. The observation of the energy spectrum of
the SN1987A $\bar{\nu}_e$'s \cite{ssb} may be used to provide very
stringent constraints on {\sl massless} neutrino mixing angles, as
seen in \fig{SN87}.  The regions to the right of the solid curves are
forbidden, those to the left are allowed. Massless neutrino mixing may
also have important implications for $r$-process nucleosynthesis in
the supernova \cite{qian}. For details see ref. \cite{massless}.

\begin{figure}[t]
\centerline{\protect\hbox{
\psfig{file=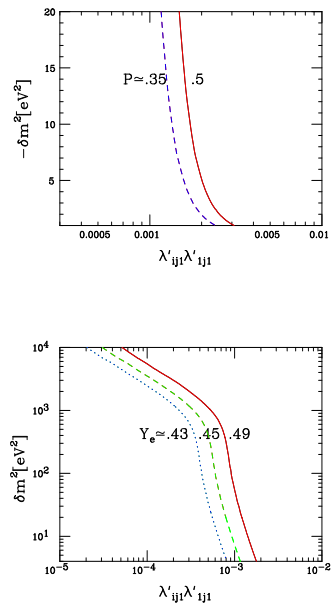,height=11.5cm,width=8cm}
}}
\vglue -2cm
\caption{Supernovae and FCNC neutrino interactions. }
\label{rpsusysnova}
\vglue -.5cm
\end{figure}

Another illustration of how supernova restricts \neu properties has
been recently considered in ref. \cite{rsusysn}. There flavour
changing neutral current (FCNC) \neu interactions were considered.
These may induce resonant massless-neutrino conversions in a dense
supernova medium, both in the massless and massive case. The
restrictions that follow from the observed $\bar\nu_e$ energy spectra
from SN1987A and the supernova $r$-process nucleosynthesis provide
constraints on supersymmetric models with $R$ parity violation, which
are much more stringent than those obtained from the laboratory. In
\fig{rpsusysnova} we display the constraints on explicit
$R$-parity-violating FCNCs in the presence of non-zero neutrino masses
in the hot dark matter eV range.  As seen from \fig{rpsusysnova} they
disfavour a leptoquark interpretation of the recent HERA anomaly.

\begin{figure}[t]
\centerline{\protect\hbox{
\psfig{file=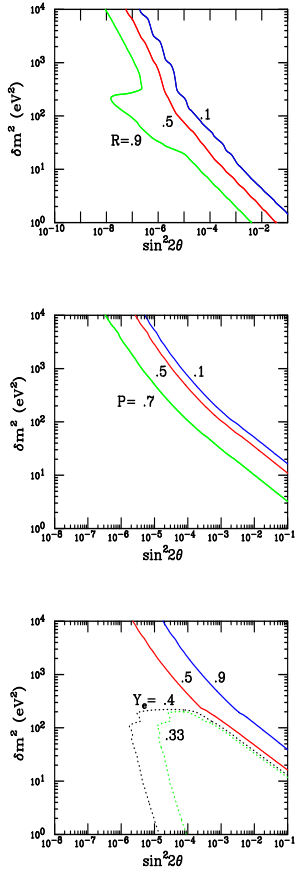,height=14cm,width=8cm}}}
\vglue -2cm
\caption{Supernovae and sterile neutrinos. }
\vglue -.8cm
\label{sterileSN}
\end{figure}

As a final example of how astrophysics can constrain \neu properties
we consider the case of resonant $\nu_e \rightarrow\nu_s$ and
$\bar{\nu}_e\rightarrow\bar{\nu}_s$ conversions in supernovae, where
$\nu_s$ is a {\it sterile} neutrino \cite{nunus}, which we assume to
be in the hot dark matter mass range.  The implications of such a
scenario for the supernova shock re-heating, the detected $\bar\nu_e$
signal from SN1987A and for the $r$-process nucleosynthesis hypothesis
have been recently analysed \cite{nunus}. In \fig{sterileSN}, taken
from \cite{nunus}, we summarize the resulting constraints on mixing
and mass difference for the $\nu_e-\nu_s$ system that follow from
these arguments.  Notice that for the case of $r$-process
nucleosynthesis there is an allowed region for which the $r$-process
nucleosynthesis can be enhanced. In fact, strictly speaking, only
SN1987A can yield real bounds on sterile neutrino parameters.

\section{HINTS FOR NEUTRINO MASSES}
\vskip .2cm

So far the only indications in favour of nonzero neutrino rest
masses have been provided by astrophysical and cosmological 
observations, with a varying degree of theoretical assumptiqons.
We now turn to these.

\subsection{Dark Matter}
\vskip .2cm

Considerations based on structure formation in the Universe have
become a popular way to argue in favour of the need of a massive
neutrino \cite{cobe2}. Indeed, by combining the observations of cosmic
background temperature anisotropies on large scales performed by the
COBE satellite \cite{cobe} with cluster-cluster correlation data
e.g. from IRAS \cite{iras} one finds that it is not possible to fit
well the data on all scales within the framework of the simplest cold
dark matter (CDM) model. The simplest way to obtain a good fit is to
postulate that there is a mixture of cold and hot components,
consisting of about 80 \% CDM with about 20 \% {\sl hot dark matter}
(HDM) and a small amount in baryons.  The best candidate for the hot
dark matter component is a massive neutrino of about 5 eV.  It has
been argued that this could be the tau neutrino, in which case one
might expect the existence of \ne $\ra$ \nt or \nm $\ra$ \nt
oscillations. Searches are now underway at CERN \cite{chorus}, with a
similar proposal at Fermilab. This mass scale is also consistent with
the hints in favour of neutrino oscillations reported by the LSND
experiment \cite{lsnd}.

\subsection{Solar Neutrinos}
\vskip .2cm

\begin{figure}[t]
\centerline{\protect\hbox{
\psfig{file=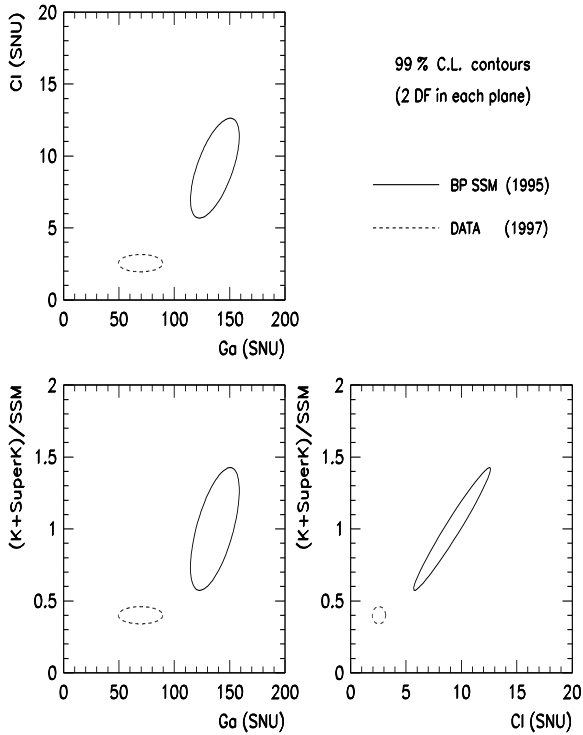,height=11cm,width=8cm}}}
\vglue -1.5cm
\caption{Solar neutrinos: theory versus data. }
\label{solardata200sk}
\end{figure}

The averaged data collected by the chlorine \cite{cl}, Kamiokande
\cite{k}, as well as by the low-energy data on pp neutrinos from the
GALLEX and SAGE experiments \cite{ga,sa} still pose a persisting
puzzle, now re-confirmed by the first 200 days of Super-Kamiokande (SK)
data. The most recent data can be summarised in \fig{solardata200sk}
where the theoretical predictions refer to the BP95 SSM prediction of
ref. \cite{SSM}.  For the gallium result we have taken the average of
the GALLEX \cite{ga} and the SAGE measurements \cite{sa}.

\begin{figure}[t]
\psfig{file=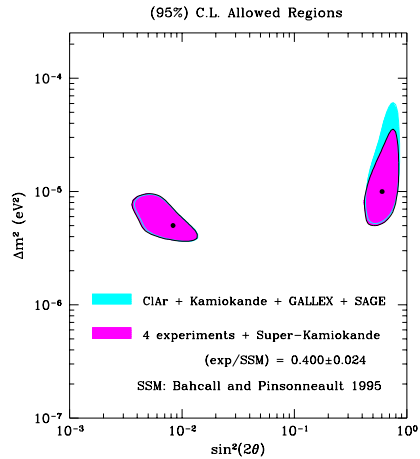,width=7cm,height=7cm}
\vglue -1.3cm
\caption{Allowed solar neutrino oscillation parameters for
active neutrino conversions.}
\vglue -0.5cm
\label{msw}
\end{figure}

The totality of the data strongly suggests that the solar neutrino
problem is real, that the simplest astrophysical solutions are ruled
out, and therefore that new physics is needed \cite{CF}. The most
attractive possibility is to assume the existence of \neu conversions
involving very small \neu masses. In the framework of the MSW effect
\cite{MSW} the required solar neutrino parameters $\Delta m^2$ and
$\sin^2 2\theta$ are determined through a $\chi^2$ fit of the
experimental data. In \fig{msw} , taken from ref. \cite{bks}, we show the
allowed two-flavour regions obtained in an updated MSW analysis of the
solar \neu data including the the recent SK 200 days data, in the BP95
model for the case of active neutrino conversions. The analysis of
spectral distortion as well as day-night effect plays an important
role in ruling out large region of parameters. Compared with
previously, the impact of the recent SK data is felt mostly in the
large mixing solution which, however, does not give as good a fit as
the small mixing solution, due mostly to the larger reduction of the
$^7$Be flux found in the later. 
The most popular alternative solutions to the solar neutrino anomaly
include the MSW sterile neutrino conversions, as well as the just-so
or vacuum oscillation solution. Recent fits have also been given
including the recent SK data \cite{bks}.

A theoretical point of direct phenomenological interest for Borexino
is the study of the possible effect of random fluctuations in the
solar matter density \cite{BalantekinLoreti}. The existence of noise
fluctuations at a few percent level is not excluded by the SSM nor by
present helioseismology studies. They may strongly affect the $^7$Be
neutrino component of the solar neutrino spectrum so that the Borexino
experiment should provide an ideal test, if sufficiently small errors
can be achieved. The potential of Borexino in "testing" the level of
solar matter density fluctuations is discussed quantitatively in
ref. \cite{noise}.
 
\subsection{Atmospheric Neutrinos}

Two water Cerenkov underground experiments, Kamiokande and IMB, and
possibly also Soudan2, have indications which support an apparent
deficit in the expected flux of atmospheric $\nu_\mu$'s relative to
that of $\nu_e$'s that would be produced from conventional decays of
$\pi$'s, $K$'s as well as secondary muon decays \cite{Barish}.
Although the predicted absolute fluxes of \neus produced by cosmic-ray
interactions in the atmosphere are uncertain at the 20\% level, their
ratios are expected to be accurate to within 5\%. While some of the
experiments, such as Frejus and NUSEX, have not found a firm evidence,
it has been argued that there may be a strong hint for an atmospheric
neutrino deficit that could be ascribed to \neu oscillations.
Kamiokande data on higher energy \neus strengthen the case for an
atmospheric \neu problem.  In ref. \cite{atm} the impact of recent
experimental results on atmospheric neutrinos from experiments such as
Superkamiokande and Soudan on the determinations of atmospheric
neutrino oscillation parameters is considered, both for the $\nu_\mu
\rightarrow \nu_\tau$ and $\nu_\mu \rightarrow \nu_e$ channels.  In
performing this re-analysis theoretical improvements in flux
calculations as well as neutrino-nucleon cross sections have been
taken into account.  The relevant oscillation parameters can be
determined from a fit and the allowed regions of parameters are found
in ref. \cite{atm}. One of the features is that the best fit value of
the $\Delta m^2$ is somewhat lower than previously obtained.

\section{RECONCILING PRESENT HINTS}
\vskip .2cm

\subsection{Almost Degenerate Neutrinos}
\vskip .2cm

The above observations from cosmology and astrophysics do seem to
suggest a theoretical puzzle. As can easily be understood just on the
basis of numerology, it seems rather difficult to reconcile the three
observations discussed above in a framework containing just the three
known \neus. The only possibility to fit these observations in a world
with just the three known neutrinos is if all of them have nearly the
same mass $\sim$ 2 eV \cite{caldwell}. This can be arranged, for
example in general seesaw models which also contain an effective
triplet vacuum expectation value \cite{LR,2227} contributing to the
light neutrino masses. This term should be added to \eq{SEESAW}.  Thus
one can construct extended seesaw models where the main contribution
to the light \neu masses ($\sim$ 2 eV) is universal, due to a suitable
horizontal symmetry, while the splittings between \ne and \nm explain
the solar \neu deficit and that between \nm and \nt explain the
atmospheric \neu anomaly \cite{DEG}.

\subsection{Four-Neutrino Models}
\vskip .2cm

A simpler alternative way to fit all the data is to add a fourth \neu
species which, from the LEP data on the invisible Z width, we know
must be of the sterile type, call it \ns. The first scheme of this
type gives mass to only one of the three neutrinos at the tree level,
keeping the other two massless \cite{OLD}.

Two basic schemes of this type that keep the sterile neutrino light
due to a special symmetry have been suggested. In addition to the
sterile \neu \ns, they invoke additional Higgs bosons beyond that of
the standard model, in order to generate radiatively the scales
required for the solar and atmospheric \neu conversions. In these
models the \ns either lies at the dark matter scale \cite{DARK92} as
illustrated in \fig{pv}
\begin{figure}[t]
\psfig{file=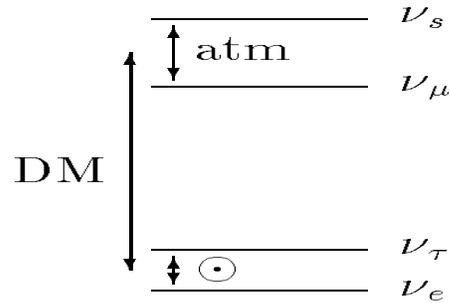,width=6cm,height=4cm}
\vglue -.5cm
\caption{{\sl "Heavy"} Sterile 4-Neutrino Model}
\label{ptv}
\vglue -.5cm
\end{figure}
or, alternatively, at the solar \neu scale \cite{DARK92B}. 
\begin{figure}
\psfig{file=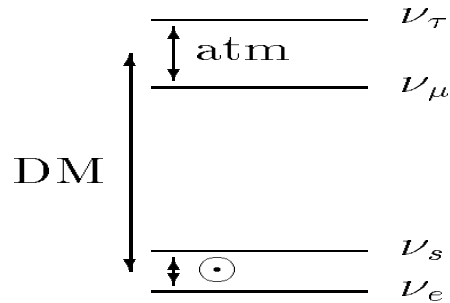,width=6cm,height=4cm}
\vglue -.5cm
\caption{{\sl Light} Sterile 4-Neutrino Model}
\label{pv}
\vglue -.5cm
\end{figure}
In the first case the atmospheric \neu puzzle is explained by \nm to
\ns oscillations, while in the second it is explained by \nm to \nt
oscillations. Correspondingly, the deficit of solar \neus is explained
in the first case by \ne to \nt oscillations, while in the second it
is explained by \ne to \ns oscillations. In both cases it is possible
to fit all observations together.  However, in the first case there is
a clash with the bounds from big-bang nucleosynthesis. In the latter
case the \ns is at the MSW scale so that nucleosynthesis limits are
satisfied. They nicely agree with the best fit points of the
atmospheric neutrino parameters from Kamiokande \cite{atm}. Moreover,
it can naturally fit the hints of neutrino oscillations of the LSND
experiment \cite{lsnd}.  Another theoretical possibility is that all
active \neus are very light, while the sterile \neu \ns is the single
\neu responsible for the dark matter \cite{DARK92D}.

\subsection{Mev Tau Neutrino}
\vskip .2cm

An MeV range tau neutrino is an interesting possibility to consider
for two reasons. First, such mass is within the range of the
detectability, for example at a tau-charm factory \cite{jj}. On the
other hand, if such neutrino decays before the matter dominance epoch,
its decay products would add energy to the radiation, thereby delaying
the time at which the matter and radiation contributions to the energy
density of the universe become equal. Such delay would allow one to
reduce the density fluctuations on the smaller scales purely within
the standard cold dark matter scenario, and could thus reconcile the
large scale fluctuations observed by COBE \cite{cobe} with the
observations such as those of IRAS \cite{iras} on the fluctuations on
smaller scales.

In ref.~\cite{JV95} a model was presented where an unstable MeV
Majorana tau \neu naturally reconciles the cosmological observations
of large and small-scale density fluctuations with the cold dark
matter model (CDM) and, simultaneously, with the data on solar and
atmospheric neutrinos discussed above. The solar \neu deficit is
explained through long wavelength, so-called {\sl just-so}
oscillations involving conversions of \ne into both \nm and a sterile
species \ns, while the atmospheric \neu data are explained through \nm
$\ra$ \ne conversions. Future long baseline \neu oscillation
experiments, as well as some reactor experiments will test this
hypothesis. The model assumes the spontaneous violation of a global
lepton number symmetry at the weak scale.  The breaking of this
symmetry generates the cosmologically required decay of the \nt with
lifetime $\tau_{\nu_\tau} \sim 10^2 - 10^4$ seconds, as well as the
masses and oscillations of the three light \neus \ne, \nm and \ns
required in order to account for the solar and atmospheric \neu
data. One can verify that the big-bang nucleosynthesis constraints
\cite{KTCS91,DI93} can be satisfied in this model.

\section{CONCLUSION}
\vskip .2cm

Although unpredicted, neutrino masses, are strongly favoured by
present models of elementary particles. On the other hand, they seem
to be required to account for present astrophysical and cosmological
observations. Neutrino mass effects could show up as spectral
distortions in many weak decays, such as nuclear $\beta$ decays and
$\pi\ell2$ decays.  Searches for $\beta \beta_{0\nu}$ decays with
enriched germanium could test the quasi-degenerate neutrino scenario
that accounts for the hot dark matter, solar and atmospheric \neu
anomalies.  Underground experiments Superkamiokande, Borexino, and
Sudbury will shed more light on the solar neutrino issue.  Oscillation
searches with long-baseline experiments both at reactors and
accelerators show good prospects for testing the regions of
oscillation parameters presently suggested by the atmospheric \neu
anomaly. Finally, new satellite experiments will test different models
of structure formation, and shed light on the possible role of
neutrinos as dark matter.

Despite all the limits from laboratory experiments, both at
accelerators and reactors, as well as the limits from cosmology and
astrophysics, there is considerable room for interesting new effects
in the neutrino sector. These cover an impressive range of energies
and could be probed in experiments performed at underground
installations as well as particle accelerators such as LEP and LHC.

\bibliographystyle{ansrt}

\end{document}